\documentclass{aa}
\usepackage{graphicx}
\usepackage{verbatim}
\usepackage{amssymb}
\usepackage{natbib}
\bibpunct{(}{)}{;}{a}{}{,}
\newcommand{\pas}{.\hskip-2pt$^{\prime\prime}$}
\def\kms{km~s$^{-1}\,$}

\newcommand{\sunp}{$_{\odot}\,$}
\titlerunning{VLBA water masers in AFGL~5142}
\begin{document}
   \title{Tracing the base of protostellar wind(s) towards the high-mass star forming region AFGL~5142:\\
VLA continuum and VLBA  H$_2$O maser observations}

  \author{C. Goddi \inst{1}\fnmsep\inst{2} 
\and L. Moscadelli \inst{1}}

   \offprints{C. Goddi,\\\email{cgoddi@ca.astro.it}}

   \institute{INAF, Osservatorio Astronomico di Cagliari, Loc. Poggio dei Pini,
 Str. 54, 09012 Capoterra (CA), Italy
   \and
      Dipartimento di Fisica, Universit{\`a} degli Studi di Cagliari,
       S.P. Monserrato-Sestu Km 0.7, I-09042 Cagliari, Italy
       }

   \date{Received ``date'' / Accepted ``date''}

   \abstract{
 We have conducted phase-reference multi-epoch observations of the 22.2~GHz  water masers    using the Very Long Baseline Array (VLBA)    and  multi-frequency study of the  continuum emission  using the  Very Large Array (VLA) towards the high-mass star forming region (SFR) AFGL~5142.
29  maser features were identified and most of them were  persistent over the four observing epochs, allowing {\it absolute} proper motions to be determined.
 The water maser emission comes from two elongated structures (indicated as Group I and Group II), with the measured  proper motions aligned along the structures' elongation axes.
 Each group consists of   two  (blue- and red-shifted) clusters of features separated by  a few hundreds and thousands  of AU respectively for Group I  and Group II.

The    maser features of Group II have both positions and     velocities aligned along a direction close to the axis of the  outflow traced by  HCO$^+$ and SiO  emission on angular scales of tens of arcsec. We predict that   the maser emission  arises from dense, shocked molecular clumps  displaced along the axis of the  molecular outflow.
 The two maser clusters of Group I are oriented on the sky along a direction forming a large angle ($\gtrapprox 60^{\circ}$) with the axis of the jet/outflow  traced  by Group II  maser features. We have detected a compact (8.4 and 22~GHz) continuum source (previously reported at 4.9 and 8.4~GHz) that falls close to the centroid of Group I masers, indicating that the source ionizing the gas is also responsible for the excitation of the  water masers.
  The kinematic analysis  indicates that the Group I masers trace outflowing rather than rotating gas, discarding the Keplerian disk scenario  proposed in a previous paper for Group I. 
Since the axis joining the two maser clusters of Group II does not cross the position of the continuum source, Group II masers might be excited by an (undetected) massive YSO, distinct from the one (pinpointed by the VLA continuum emission)   responsible for the excitation of the Group I masers. 

Our results give support to models of accretion and jet ejection related to the formation of high-mass stars.

%   \keywords{ Masers --  Stars: formation -- ISM: kinematics and dynamics -- ISM: jets and outflows -- Accretion, accretion disks -- Radio continuum: ISM } 
               }

  \maketitle
%
%____________________________________________________________________________________________________________
\section{Introduction}
A widely accepted, observationally supported, ``standard'' theory exists for the formation of low-mass, solar-type stars, involving inside-out collapse of molecular cloud fragments and formation of  accretion disks (e.g. \citealt{Shu87}). For high-mass  ($\gtrapprox 10$ M\sunp) stars, on the contrary, further effort (from both a theoretical and observational point of view) is required to improve our understanding of their formation process. A fundamental question is whether the mode of high-mass formation is a scaled-up version of the low-mass forming mode or weather it involves different physical processes. This question arises because stars  with masses larger than 8 M$_{\odot}$   contract to the main sequence and ignite while still accreting matter from the infalling envelope. Then, the "standard" theory predicts that, for a spherically symmetrical  collapse, the radiation pressure  produced by the massive YSO enormous luminosity would be sufficient to inhibit any further accretion. That leads to the paradox that massive stars ($M > 8$ M$_{\odot}$) might never form, in disagreement with the observed upper limits of  hundreds of  solar masses. In order to circumvent this dilemma, it has been proposed that  high-mass stars might build up via merging of lower mass ($M < 8$ M$_{\odot}$) YSOs (coalescence model -  e.g. \citealt{Bon02}), or, alternatively, by accreting material from the infalling envelope via a thin disk, as it occurs in low-mass stars (accretion model - e.g.  \citealt{Yor02}). In the disk scenario, in order to allow the gas of the disk to accrete onto the stellar surface, the removal of the excess angular momentum is thought to be "somehow" associated to the ejection of a powerful bipolar outflow, collimated along the disk-axis. Then the accretion paradigm would predict the occurrence of disk/outflow systems, whereas the alternative coalescence scenario would imply the destruction of ordered kinematical structures surrounding the merging protostars. 

According to  widely accepted theories, (proto)stellar jets may be launched by the combined action of magnetic and centrifugal forces along the magnetic field lines threading the YSO's envelope. 
 The  magnetohydrodynamical (MHD) models  proposed to explain the origin of (proto)stellar jets can be divided into two categories, depending on the place where the jet is supposed to originate: 1) {\it X-winds}, which are magnetized stellar winds driven by the {\it stellar} magnetic field,  arising at the star/disk boundary  (see \citealt{Shu00} for a review); 2)  {\it disk-winds}, where the jet stems from the surface of a Keplerian disk threaded by the {\it circumstellar} magnetic field  (\citealt{Kon00} and references therein). 
 
 A possible way, from an observational point of view, to distinguish between  alternative (accretion and jet ejection) models in massive star formation theories is studying the kinematics of the gas in the proximity  ($<100$~AU) of the YSO, corresponding to the "root" of the putative jet/disk system.
%MASER
An ideal diagnostic tool  is provided by Very Long Baseline Interferometry (VLBI) observations at centimeter (cm) wavelengths of maser transitions of molecular species (such as  OH, H$_{2}$O, CH$_{3}$OH).
In particular, for fast-moving 22.2~GHz water masers, multi-epoch VLBI observations, reaching  angular resolutions of  $\simeq $0.5~mas (corresponding to $\simeq$1~AU at a distance of 2~kpc), permit accurate proper motions to be determined with time baselines as short as a few months. Combining the measured tangential velocities with the radial velocities derived via the Doppler effect, allows  one to obtain the 3D velocity distribution of the masing gas. 

Previous VLBI observations have shown that 22.2~GHz water masers  are preferentially associated with collimated (jets) or wide-angle flows of gas   at the base of larger-scale molecular outflows (W3~IRS~5, \citealt{Ima00};  W75N-VLA1, \citealt{Tor03}; Sh~2-255 IR, IRAS~23139+5939, WB89-234, and OMC2, \citealt{God05};  IRAS~20126+4104, \citealt{Mos05}). In  a few objects (NGC 2071 IRS~1/3, \citealt{Set02}; AFGL~5142, \citealt{God04} - hereafter Paper I),   accretion disks have been proposed to explain linear clusters (size $\sim$10-100~AU) of H$_2$O masers detected through VLBI observations.

The 22.2~GHz H$_2$O masers in AFGL~5142 (for a detailed description of the source see Paper I) were  observed with the European VLBI Network (EVN) at four epochs (from October 1996 to November 1997) and found to be distributed across two  structures (indicated as ``Group I'' and ``Group II''). Such structures  have been interpreted as being possibly associated with a disk/jet system, indicating that AFGL~5142 may be an ideal candidate to test the formation theories of high-mass stars. However, the proposed interpretation  needs more stringent observational constraints,  the multiepoch EVN observations  suffering  two major drawbacks: \ 1)~too long  time separation  between consecutive epochs ($\geq$ 3~months, more than the average life time of the maser features); \ 2)~limited sensitivity  (average detection threshold of $\sim$0.3~Jy~beam$^{-1}$). These limitations prevented us to detect  and measure the proper motions of the weakest and less-longeval maser features. 

Recently, we performed follow-up observations of the 22~GHz H$_2$O masers in AFGL~5142 using the Very Long Baseline Array (VLBA), taking advantage of both  high sensitivity and an optimized  time separation  between  consecutive epochs. The VLBA observations were performed in phase-reference mode to derive the absolute positions and proper motions of the water maser features.

We have also performed multi-frequency VLA-B continuum observations of AFGL~5142, aiming to precisely locate the position of the exciting  YSO(s), to put  constraints on the nature of the previously observed radio continuum emission   \citep{Car90,McC91,Tor92,Hun95,Hun99,Car99},  and, finally, to investigate the relationship between the water masers and the continuum source.

Section~\ref{obs} describes our VLA and multi-epoch VLBA observations and gives technical details on the data analysis. Section~\ref{res} compares our  results with previous interferometrical observations of the region.
 In Section~\ref{dis}, plausible kinematic models   for interpreting the measured positions and velocities of the maser features are investigated. Conclusions are drawn in Section~\ref{con}.

%__________________________________________________________________
\section{Observations and data analysis}
\label{obs}
\subsection {VLA Observations}

The continuum observations were performed using the VLA in its B configuration on  March 1 and 11 2005, for a total observing time of 4 hours. We observed a bandwidth of 50 MHz centered at a central frequency of 8.4601, 22.4601, and 43.3399~GHz ({\it X}, {\it K}, and {\it Q}-bands, respectively).
The data were edited, calibrated, and imaged using the NRAO's Astronomical Image Processing System (AIPS) package. Absolute amplitude calibration was obtained from observations of 3C147 (with flux density of 4.8, 1.8, and 0.9~Jy for {\it X}, {\it K}, and {\it Q}-band, respectively). In each band, the observations were performed using the ``fast switching'' method, where short (420, 180, 60~s) scans of AFGL~5142 were alternated with 60 s  scans of the nearby phase calibrator J05181+33062.
The error in the absolute calibration is estimated to be within 10\% in {\it X}, 20\% in {\it K} and  30\% in {\it Q}-band.
%Q-band observations are more affected by variations in the weather conditions and gain variations with telescope elevation than lower frequency observations.
Table~\ref{VLA} reports the  main  observational parameters (restoring beam  FWHM sizes and position angle, and the map RMS noise level) for the three observing bands.
% were (0\pas88, 0\pas82, $71^{\circ}$), (0\pas38, 0\pas38, 0$^{\circ}$), and (0\pas17,0\pas15, $31^{\circ}$), whereas the achieved RMS noise levels were 26, 52, and 270 $\mu$Jy beam$^{-1}$, for the  {\it X}, {\it K}, and {\it Q}-band maps, respectively.
%
\begin{table*}
\centering
\caption {Radio continuum observations.}
\begin{tabular}{ccccccccc}
\hline\hline
Band & Beam size & Beam P.A. & $\sigma$  & $ \alpha(J2000)$ & $\delta(J2000)$ & $S_{\nu}$  &FWHM Size & P.A.\\
 & ($''$) & ($^{\circ}$) & ($\mu$Jy~beam$^{-1}$) & (h m s)  & ($^\circ \ '\ ''$)& (mJy)  & (arcsec)& ($^{\circ}$)  \\
\hline
X & $0.88 \times 0.82$ & 71 & 26  & 05 30 48.019  & 33 47 54.6&   $0.75\pm 0.08^a$ &-&- \\
K & $0.38 \times 0.38$ & 0 & 52  & 05 30 48.020 & 33 47 54.5&  $0.7\pm 0.1$ & $0.35\times0.27^b$ & 130 \\
Q & $0.17 \times 0.15$ & 31 & 270 & -&-& $< 0.9^c$ &-&-\\ 
\hline
\end{tabular}
\begin{flushleft}
{\footnotesize  Note.-- For each observing band, Cols.~2 and 3 give respectively the FWHM sizes along the  major and minor axes, and the position angle of the restoring beam; Col.~4   the RMS noise level on the maps; Cols.~5 and 6 the R.A. and DEC absolute position of the intensity peak; Col.~7 the  integrated flux density of the maps;  Cols.~8 and 9  the deconvolved FWHM sizes and the position angle of the detected sources.

$^a$  The absolute flux density uncertainty  is estimated to be  $ \leq  10\%$ at {\it X}-band, $\leq 20\%$ at {\it K}-band, $\leq 30\%$ at {\it Q}-band. 

$^b$ The  source is compact at all observed frequencies, with the exception of the {\it K}-band, for which the deconvolving routine (SAD) produced FWHM sizes comparable with the observing beam.

$^c$ At 43.3~GHz the flux density upper limit corresponds to $3\sigma$. 
}
\end{flushleft}
\label{VLA}
\end{table*}

\subsection {VLBA Observations}

\subsubsection {VLBA runs}
The observations  in the \(6_{16}-5_{23}\) H$_2$O maser line at 22.2~GHz  were conducted using the VLBA at four epochs (16 October and 22 November 2003, 1 January and 8 February 2004), each epoch lasting for 12 hours.
The  observations were performed in phase-reference mode, alternating scans on   the maser source, AFGL~5142, and the phase-reference source, J0518+3306, with a  switching  cycle of 70~s.
 J05181+33062 is separated 2.6$^{\circ}$  from AFGL~5142, belongs to the VLBA calibrator catalog, and has very accurate coordinates (R.A. and DEC uncertainties $< 1 $~mas). Interlapsed every $\approx$80~min, 3--min scans on several continuum sources (0528+134, 0552+398, 0333+321, 0642+449) were observed  for calibration purposes.  The total on-source integration time for the maser target was about 5.5 hours.

Both circular polarizations were recorded using a 16~MHz bandwidth centered on the LSR velocity of --4.8~km~s$^{-1}$. The data were correlated with the VLBA FX correlator in Socorro (New Mexico) with an integration period of 1~s. The correlator used  1024 spectral channels corresponding to a channel separation of 0.2~km~s$^{-1}$.

\subsubsection{Data reduction}
Data reduction was performed using the  AIPS package, following the standard procedure for VLBI line data.
Total power spectra of the continuum calibrators were used to derive the bandpass response of each antenna. The amplitude calibration was performed using the information on the system temperature and the gain curve of each antenna. 

For each observing epoch, a single scan of a strong calibrator was used to derive the 
instrumental (time-independent) single-band delay  and the phase offset between the 
two polarizations. %(caused by passage of the signal through the electronics of the 
%video converter units).
After removing the instrumental errors, all calibrator scans were fringe-fitted to determine the residual (time-dependent) delay and the fringe rate. 
%Since the maser source  was significantly stronger than the phase-reference source, we applied the {\em reverse} phase referencing technique.
The corrections derived from calibrators were applied to the strongest maser component, chosen to refer the visibilities of all the other maser emission channels.
In  each observing epoch the strongest maser component emitted at the same velocity, $V_{\rm LSR}$ = --4.9~km~s$^{-1}$, and exhibited a simple spatial structure consisting of a single, almost unresolved spot.
The visibilities of the reference channel were fringe-fitted to find the 
residual fringe rate produced both by differences in atmospheric fluctuations between 
the calibrators and the maser, and by errors in the model used at the correlator. 
After correcting for the residual fringe rate, the visibilities of the reference channel were 
self-calibrated to remove any possible effect of spatial structure. Finally, the corrections derived 
from the reference channel were applied to data of all spectral channels.

Spectral channel maps were produced extending over a sky area of $(\Delta \alpha \ cos\delta \times \Delta \delta) \ 6''\times 6''$ and covering the whole velocity range where signal was visible in the total-power spectra (from  $-$10 to 5~km~s$^{-1}$). At each epoch, the maser emission centres are found to be distributed within an area of $2''\times $1\pas5.
The CLEAN beam was an elliptical Gaussian with a  FWHM size of $ 0.7 \times 0.4 $~mas. In each observing epoch, the RMS noise level on the channel maps, $\sigma$, varied over a similar range of values, 4--40~mJy~beam$^{-1}$.

\begin{table*}
\centering
\caption {Offsets of the reference feature from the correlated position: $\alpha(J2000) = 05^h 30^m 48$\pas033,  $\delta(J2000) = 33^{\circ} 47' 54$\pas6; Col.~1 reports the observing epoch; Cols.~2 and 3 respectively the R.A. and DEC apparent  offsets, as derived from the phase-reference calibrated maps;  Cols.~4 and 5 the corrections for the combined action of Galactic, Sun and earth motion; Cols.~6 and 7 the corrected (R.A. and DEC) offsets.}
\begin{tabular}{ccccccccc}
\hline\hline
Epoch & $\Delta \alpha_1$ & $\Delta \delta_1$ & $\Delta \alpha_2$ & $\Delta \delta_2$ & $\Delta \alpha_f$ & $\Delta \delta_f$\\
& (mas) & (mas) & (mas) & (mas) & (mas) & (mas)\\
\hline
16 Oct 03 & --83.1 (0.4)& --244.7 (0.2) & --1.2 & 1.0 & --84.3 (0.4) & --243.7 (0.2)\\
22 Nov 03 & --83.4 (0.1)& --244.6 (0.1)& --0.9& 1.2 & --84.3 (0.1) & --243.4 (0.1)\\
1 Jan 04 &--83.7 (0.3) & -245.3 (0.2) & --0.6 & 1.4 & --84.3 (0.3) & --243.9 (0.2)\\
8 Feb 04 & --84.0 (0.3)& -245.5 (0.2)& --0.3 & 1.7 & --84.3 (0.3) & --243.8 (0.2)\\
\hline
\end{tabular}
\label{off}
\end{table*}

\subsubsection{Absolute position determination}
\label{abs}
Auto-referencing the visibilities phases of the maser data   causes the  loss of the absolute position information.
Since in our case the water maser source is significantly stronger than the phase reference source, J0518+3306, we first attempted to determine the maser absolute position using the reverse phase referencing technique.
All the phase calibration (fringe-fitting + self-calibration) solutions obtained working on the reference  maser component were applied to the J05181+33062 data.
So doing, the position of the phase-reference source is shifted by a vector  equal in absolute value but with opposite sign respect to the offset of the maser reference spot from its  correlated position.
Finally, the maser-referenced visibilities of J05181+33062 were Fourier-transformed, but the resulting map produced only  a marginal detection (with a peak flux density of $\sim 10$~mJy).

Then, we experimented the alternative way of direct phase-referencing, fringe-fitting the scans of J05181+33062  and applying the obtained phase calibration solutions to the maser data. The image constructed by Fourier transforming the   visibilities of the maser source, calibrated with the phase-reference continuum solutions,
 presented a signal-to-noise ratio sufficiently high to allow us to determine the offset of the reference maser (from the position used at the correlator) with an accuracy of tenths of milliarcsec. 
For the first observing epoch, the derived  positional offset is: $ \Delta\alpha_1 = -$0\pas0831 $\pm$ 0\pas00035,  $\Delta\delta_1 = -$0\pas24469 $\pm$ 0\pas0002. The uncertainty on the positional offset is estimated by the ratio of the FWHM beam size to the dynamic range of the map.
In order to obtain the peculiar motion of the maser source, the offsets derived at each epoch have been   corrected for  systematic effects  associated to Galactic rotation, annual parallax and solar motion. The corrected offsets of the reference feature  are reported in Table~\ref{off}.

\begin{figure*}
\centering
\includegraphics[width = 17cm]{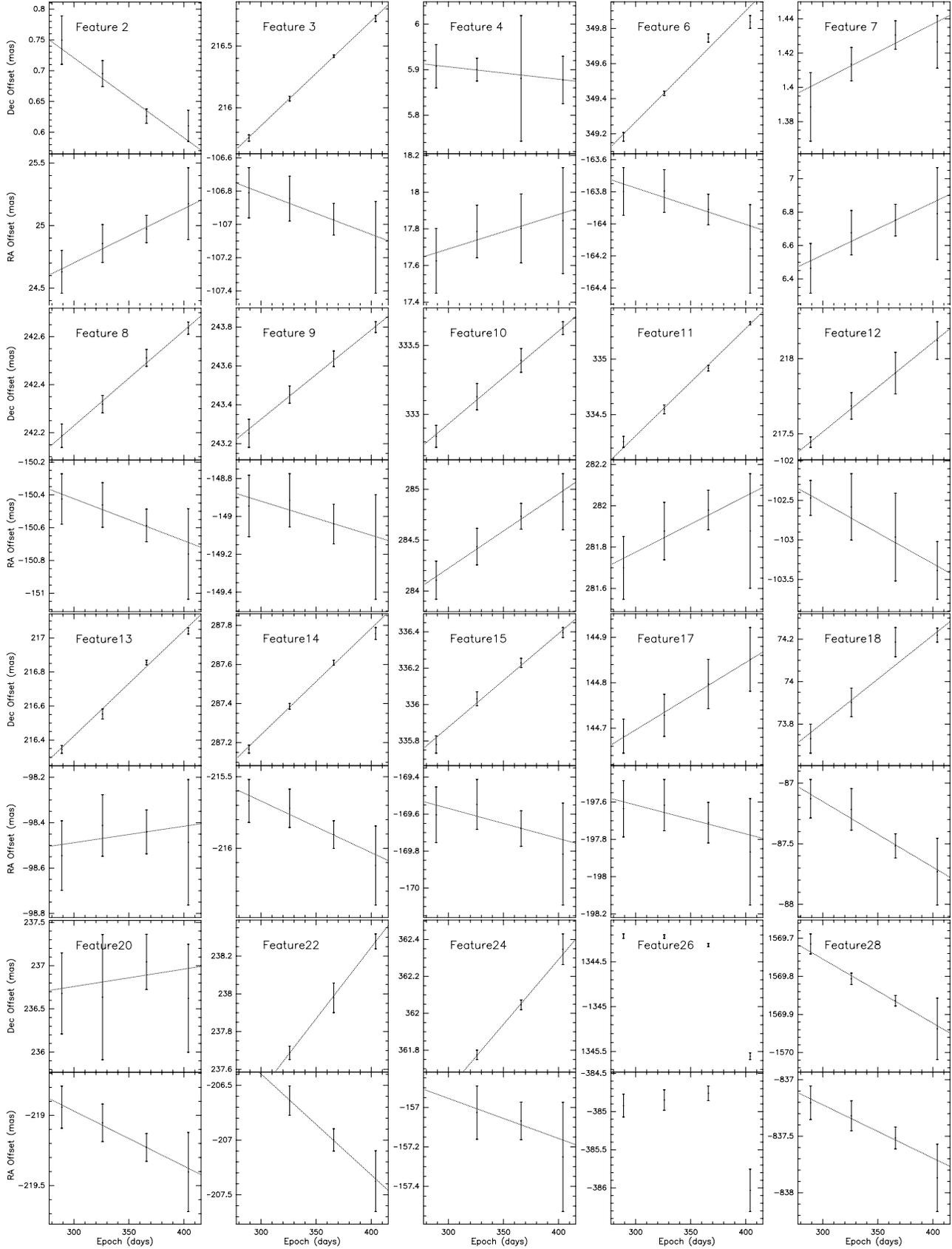}
\caption[Measured relative proper motions of the {\em VLBA} H$_2$O maser features in AFGL~5142.]{Measured {\em relative} proper motions of the VLBA H$_2$O maser features in AFGL~5142.
For each of the time persistent features, the top and the bottom panels report the time variation of, respectively, the DEC and the R.A. offsets (relative to the reference feature "1"). 
In each panel, the dotted line shows the proper motion calculated by the (error-weighted) linear least-squares fit of the positional offsets with time.  No proper motion is derived for the  feature with label number "26", since it does not show a linear  variation of positional offsets with time.
}
\label{prmaf_vlba}
\end{figure*}
\subsubsection{Identification of maser emission}
Each  channel map of the maser source  was searched for emission above a  detection threshold  taken equal to the absolute value of the minimum in the map  (varying across the channels in the range 5--7~$\sigma$).
The detected maser spots were fitted with two-dimensional elliptical Gaussians, determining  position, flux density, and FWHM size of the emission.
Consistently to Paper I, the term of ``feature'' is used to indicate a collection of spectrally and spatially contiguous maser spots.
A maser feature is considered real if it is detected in at least two contiguous channels (i.e., line-width $> 0.4$~\kms), with a position shift of the intensity peak from channel to channel smaller than the FWHM size.
The VLBA observations provided an improvement in sensitivity of about an order of magnitude  respect to the previous EVN observations, lowering the detection threshold to $\sim$ 30~mJy~beam$^{-1}$ and allowing the detection of a much larger number of  maser spots and features.
In each observing epoch, about 270 maser spots were identified, grouped  into 29 distinct maser features. 
Table~\ref{af_vlba} gives the parameters of the identified features.
 The absolute (R.A. and DEC) positional offsets of a given feature, evaluated with respect to the maser position used at the correlator ($\alpha(J2000) = 05^h 30^m 48$\pas033,  $\delta(J2000) = 33^{\circ} 47' 54$\pas6), are calculated with a two step procedure.
First,   relative positions (respect to the reference maser spot) are derived  from the (error-weighted) mean of  the  maser spot positions contributing to the feature emission. Then, at each epoch the relative positions are combined with  the  absolute positional  offsets of the reference spot (reported in Table~\ref{off}).
For each feature, the  absolute positional errors  are estimated taking the sum of  the feature relative positional uncertainties (evaluated by  the weighted standard deviation of the spot positions) with the absolute   positional uncertainties of the reference spot.

%____________________________________________________________________________________________________
\subsubsection{Proper motion measurements}
\label{prop_mot}
For the features persistent over three or four epochs, the proper motions have been calculated  performing a (error-weighted) linear least-squares fit of the positional offsets with time.
%Features observed at only two epochs were not considered since [their deviations from the straight line motion were not estimable][the criterion adopted to evaluate the time persistence is that features are considered to be correspondent from epoch to epoch only in the case that their positions align along a straight line (``constant velocity'' approximation)].
For the VLBA observations the  time separation between consecutive epochs  was shorter than for the previous EVN observations  ($\sim$1~month vs 3--6~months) and that allowed to measure the proper motions of the most variable features, increasing significantly the number of measured proper motions  (23 vs 2 of previous EVN observations, considering only features persistent over $\geq 3$ epochs).
 For a representative sample of persistent features, Fig.~\ref{prmaf_vlba} shows the time variation of the R.A. and DEC ({\it relative}) offsets and the best linear fit giving the proper motion.
Among the features observed at three or more epochs,  the proper motions are derived  only for those moving in a straight line at constant velocity (within the positional errors).
Adopting such  a criterion, no proper motion is derived for the  feature with label number "26", although detected at all the four observing epochs.%NN SI MUOVE DI MOTO COSTANTE: PRIMA {\`E} ACCEL. E BLOCC. DA MATERIALE DENSO, POI RIESCE A SUPERARLO

  The derived absolute proper motions are reported in  Table~\ref{af_vlba}. The  uncertainties of the absolute proper motions are the formal errors of the linear least-squares fits and are generally much larger than those associated with the relative proper motions, owing to the relatively large error of the absolute position of the reference spot. 
The phase-reference calibrator is too weak and too much detached from  the maser target to allow an effective removal of the atmospheric noise affecting  the maser  visibilities.

\begin{table*}
\centering
\caption{Parameters of maser features detected with the VLBA in AFGL~5142.}
\begin{tabular}{ccccccccccc}
& & & & & & &  & & & \\
\hline\hline
\multicolumn{1}{c}{} & \multicolumn{1}{c}{Feature} & \multicolumn{1}{c}{$V_{\rm LSR}$} &
\multicolumn{1}{c}{$F_{\rm int}$} &  &
\multicolumn{1}{c}{$\Delta \alpha$} & \multicolumn{1}{c}{$\Delta \delta$} &
& \multicolumn{1}{c}{$V_{\rm x}$} & \multicolumn{1}{c}{$V_{\rm y}$} &
\multicolumn{1}{c}{$V_{\rm mod}$} \\
\multicolumn{1}{c}{} & & \multicolumn{1}{c}{(km s$^{-1}$)} &
\multicolumn{1}{c}{(Jy)} &   &
   \multicolumn{1}{c}{(mas)} & \multicolumn{1}{c}{(mas)} &
   & \multicolumn{1}{c}{(km s$^{-1}$)} & \multicolumn{1}{c}{(km s$^{-1}$)} &
   \multicolumn{1}{c}{((km s$^{-1}$)} \\
\hline
& 1& --5.0 & 33.2 & & --84.3 (0.4) & --243.7 (0.2)& &0.8 (11)& --11 (7) & 11 (7) \\
& 2&--5.7 &7.1 & &--59.6 (0.4) & --242.9 (0.2)&  & 14 (14)& -15 (7) & 20 (11)\\
&3 &--6.3 &5.3 & &--191.1 (0.4) &--279.2 (0.2) &  &--9 (13) &16 (7) & 18 (9) \\
&4 &--7.8 &1.1 &&--66.6 (0.4) &--237.8 (0.2) & &5(14)  &--11 (7) &12 (9) \\
& 5&--4.4 &0.1 & &--57.9 (0.3) &--242.1 (0.2) &  & & & \\
&6 &--3.1 &0.6 & &--248.1 (0.4) &105.5 (0.2) &  &--9 (13) &8 (7)&12 (11) \\
&7 &--2.7 &1.1& &--77.8 (0.4) & 242.3 (0.2)& &8(13)  &--10 (7) &13 (10)  \\
&8 &0.1 &0.5 & &--234.7  (0.4) &0.0 (0.2) &  &--9 (13) &2 (7) &9 (13) \\
&9 &--0.2 &0.6 & &--233.2  (0.4) &--0.4  (0.2)&  &--6 (13) &4 (7) &8 (12) \\
Group I& 12&--1.4 & 0.5& &--186.7  (0.4)&--26.2 (0.2) &  &--26 (16) &9 (8) &28 (16) \\
&13 &--2.2 &0.3 & &--182.8 (0.4)&--27.4  (0.2)&  &0.9 (13) &9 (7) &9 (7) \\
&14 &--1.3 &5.2 & &--299.9 (0.4) &43.5  (0.2)&  &--13 (13) &5 (7)& 14 (13)\\
&15 &--1.7 & 0.7& &--253.9 (0.4)& 92.1 (0.2)&  &--7 (13) &6 (7)&9 (11)\\
&16 &--1.9 &0.3 & &--290.0 (0.4)& --24.0 (0.2)&  & & & \\
&17 &--1.7 &0.5 & &--281.9 (0.4)&--99.0  (0.2)&  &--6 (14) &--5 (7) &8 (11) \\
&18 &4.6 &0.9 & &--171.4 (0.4)& --169.9 (0.2)&  &--18 (14) &4 (7) &18 (13) \\
&19 &1.9 &0.5 & &--222.4 (0.5) &4.0  (0.2)&  &--0.3 (16) &11 (8) &11 (8) \\
&20 &--2.0 &0.1 & &--303.2 (0.4)&7.0 (0.5)&  &--12 (13) & --2 (20)&13 (14) \\
&21 &--1.3&0.2 &&--184.5 (0.4) &--26  (0.2)&  &--18 (13) &16 (7) &24 (11) \\
&22 & 0&0.3 & &--291.0 (0.2) &--5.8 (0.1)& &--26 (17)&5 (9) &27 (17) \\
&23 &1.8 &0.1 & &--290.1  (0.3)& --19.96 (0.2)&  & & & \\
&24 &1.3 &0.1 & &--241.4 (0.2)& 118.3 (0.1)&  &--5.0 (17) &3 (9) &6 (15) \\
&25 &--0.6 &0.1 & &--240.6  (0.3)& 118.9  (0.2)&  & & & \\
& & & & & & &  & & & \\
&10 &--0.1 &0.7 & &199.8 (0.4)&89.2  (0.2)&  &22 (14) &11 (7)&24 (13) \\
&11 &0.9 &0.1 & &197.4 (0.4)&90.6  (0.2) &  &6 (13) &19 (7)&20 (8)\\
Group II &26& --7.4& 6.6& &--469.2 (0.4)& --1587.9 (0.2)& & && \\
&27&4.1 &0.5 & &--450.0 (0.2)& --1481.0 (0.1)&  & & & \\
& 28&--5.7 &4.9 & &--921.5 (0.4)&--1813.4 (0.2) &  &--17 (14) & --16 (7)&24 (11) \\
& 29& -4.2&0.7 & &--921.6 (0.4)& --1812.9 (0.2)&  &--27 (13)&--18 (7)&32 (12) \\
& & & & & & &  & & & \\
 \hline
\end{tabular}
\begin{flushleft}
{ \footnotesize  Note.-- For each identified feature, Col.~1 gives the  label number; Cols.~2 and ~3  the line-of-sight velocity  and  the integrated flux density   of the highest-intensity channel; Cols.~4 and ~5  the {\it absolute} positional (R.A. and DEC) offsets (and, within brackets, the associated errors),  evaluated with respect to the maser position used at the correlator ($\alpha(J2000) = 05^h 30^m 48$\pas033,  $\delta(J2000) = 33^{\circ} 47' 54$\pas6); Cols.~6, ~7 and ~8  the projected components along the RA and DEC axes and the absolute value of the derived proper motions (and, within brackets, the associated errors).
}
\end{flushleft}
\label{af_vlba}
\end{table*}
\begin{figure}
\centering
\includegraphics[width=8cm]{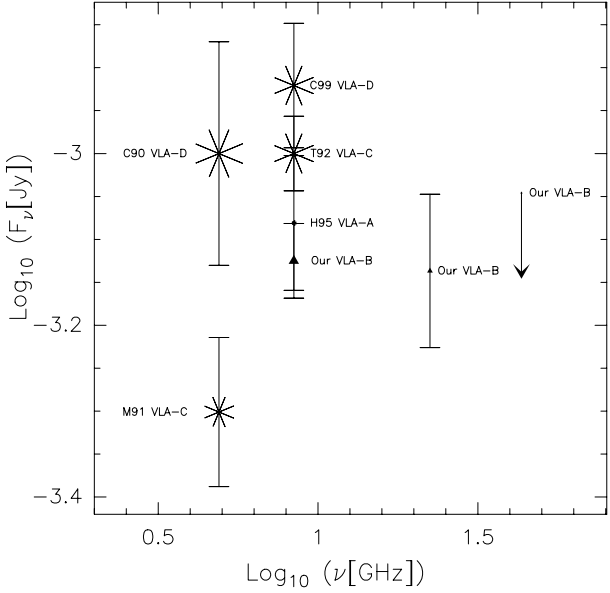}
\caption{ The radio spectral energy distribution of AFGL~5142. {\it Filled triangles} are our  VLA-B   measurements of the flux density at 8.4 (0.75~mJy with a  $\approx$~0\pas9 beam), 22.4 (0.7~mJy with a  $\approx$~0\pas4 beam), and 43.3~GHz ($\leq 0.9$~mJy with a  $\approx$~0\pas2 beam; the upper limit is indicated by the downward arrow). The {\it asterisks} are previous flux  density measurements  at 4.9~GHz (1.0~mJy  with a $\approx$~12\pas5~beam using VLA-D -- \citealt{Car90} (C90) --  and 0.5~mJy   with a $\approx4''$~beam using VLA-C -- \citealt{McC91} (M91)) and 8.4~GHz (0.83~mJy   with a $\approx$~0\pas3 beam using VLA-A -- \citealt{Hun95} (H95);  1.0~mJy  with a $\approx$~7\pas7 beam using VLA-C -- \citealt{Tor92} (T92);   1.2~mJy with a $\approx$~7\pas5 beam using VLA-D -- \citealt{Car99} (C99)). The size of the symbols scales  logarithmically with the FWHM size of the beam.
}
\label{spec_ind}
\end{figure}
%
%______________________________________________________________
\section{Observational results}
\label{res}

\subsection{Radio continuum emission}
\label{radio}
The VLA-B,  {\it X} and {\it K}-band continuum images of AFGL~5142, above a detection threshold of  $5 \sigma$, show a single,  compact  source coinciding (within the positional errors) with the previously detected emission at 4.9~GHz \citep{Car90,McC91} and 8.4~GHz \citep{Tor92,Hun95,Hun99,Car99}.
No emission was detected at 43~GHz  and only a 3$\sigma$ upper limit to the source flux density could be derived.
Table~\ref{VLA} gives the measured positions and flux densities for the three observing bands.
At K-band,  the emission appears slightly resolved, with a  deconvolved elliptical Gaussian shape having a FWHM size (estimated using the AIPS task SAD) of 0\pas35$\times$0\pas27 (corresponding to a diameter of $\sim  600$~AU), and elongated along a southeast-northwest (SE-NW) direction.

%Using the flux densities we have derived at {\it X} and {\it K}-bands plus the value ($0.5 \pm 0.1$~mJy) measured by \citet{Mcc91} in {\it C}-band, the spectral index of the continuum emission between 5 and 22~GHz is $\alpha_r = 0.18\pm 0.18$. Considering also the 43~GHz upper limit the result does not change, $\alpha_r = 0.18\pm 0.14$ (Fig.~\ref{spec_ind}).

 Previous  observations at 8.4~GHz  continuum, conducted   using the VLA in A \citep{Hun95},  B \citep{Hun99}, C \citep{Tor92}, and D \citep{Car99} configurations, all  show a {\it compact} continuum emission, with the exception of \citet{Tor92}, who presented marginal indication for the emission being extended   at scales of $\sim 20''$\footnote{In the \citet{Tor92} map,   the presence of extended emission  is suggested only by the weakest contours, below the  $3\sigma $ level.} (see Fig. 3 of \citealt{Tor92}).
Fig.~\ref{spec_ind} reports our and previous  continuum observations towards AFGL~5142. At 8.4~GHz we have reported only our VLA-B measurement, not considering the previous value of  \citet{Hun99}, which is much more uncertain\footnote{The  8.4~GHz  observations were conducted by \citet{Hun99} in snapshot mode, with a few minutes of integration, whereas our  8.4~GHz  observations last for about one hour, allowing a significant sensitivity improvement.}.
One notes that, at both  4.9 and 8.4~GHz, the integrated flux density increases with the  FWHM beam size of the observation, suggesting the presence of an extended component with size $>4''$ (the VLA-C beam). Going from  4.9 to 8.4~GHz the VLA-C flux density increases by a factor of 2 (less than a factor of  3, as expected for optically thick emission), whereas the VLA-D flux  density remains substantially constant. That is consistent with the presence of a compact core of emission at 4.9~GHz, that becomes optically thin at 8.4~GHz. Accordingly, our VLA-B measurements indicate that the integrated flux density remains constant or slightly decreases at frequencies larger than 8.4~GHz. Assuming an {\it average} optical depth $\approx 1$ at  8.4~GHz for the emission inside the VLA-B beam ($\approx$0\pas9), one finds that the{\it beam-averaged} value for the density of the ionized gas is $2 \ \times \ 10^5$~cm$^{-3}$.

Basing on the extended nature of the radio continuum and the derived high value for the density of the ionized gas, the most likely interpretation is that of a young HII region.
To account for  the measured radio flux density, an ionising flux by a ZAMS star with spectral type B2 or earlier is required, corresponding to an expected value of stellar mass  $\geq 10~M_{\odot}$ \citep{Vac96, Pal02}.

%____________________________________________________________________________________________________________________
\subsection{22.2~GHz water maser emission}

Fig.~\ref{prmot} compares our VLA and VLBA results with previous interferometric observations.
Top right-hand panels show the high-velocity molecular outflows seen in HCO$^{+}$ (1 $\to$ 0)  and SiO ($v = 0$, 2 $\to$ 1) with OVRO \citep{Hun99}.
The 22.2~GHz water masers  in AFGL~5142 (prior of our EVN observations) were observed with the VLA at two epochs (1992, \citealt{Hun95}; 1998, \citealt{Hun99}), and found  within a few arcseconds from a compact, thermal continuum source (observed both with the VLA at 8.4~GHz and with OVRO at 88~GHz), located at the center of the SiO-HCO$^+$ outflow (Fig.~\ref{prmot},  upper left-hand panel).
The lower panel of Fig.~\ref{prmot}  shows the positions and the velocities of the H$_2$O maser features as derived by our multi-epoch VLBA observations, overlaid on our VLA map of the 22~GHz continuum emission.

 Basing on the spatial distribution and the proper motion orientation of the detected  maser features, the water maser emission appears to trace two elongated structures, oriented at quite different P.A. ($\Delta$P.A.$ \gtrsim  60^{\circ}$).
Using the same nomenclature  of Paper I, we refer to such structures as Group I and Group II. In Table~\ref{af_vlba},   we report the detected maser features separately for each Group.

Masers of Group I are found closer (within $\approx$300~mas) to the 22~GHz  continuum emission peak.
It is worth noting that, basing on the accurate absolute positions we have determined for the water masers and the thermal radio source,  the 22~GHz  continuum peak falls right at the center of the maser distribution of Group I (Fig.~\ref{prmot}, lower panel). The features detected towards SE of the continuum source have LSR velocities  blue-shifted (respect to the LSR velocity of the region) whereas those detected towards NW are red-shifted. The measured transverse velocities of the blue and red-shifted features  have similar amplitudes  (in the range 6--27~\kms) but opposite versus of motion. 

Maser features of Group II are concentrated in two clusters, having a sky-projected distance of  $\sim$ 1700~mas  (corresponding to $\sim$ 3500~AU at a distance of 1.8~kpc). The line connecting the two clusters is oriented  along a northeast-southwest (NE-SW) direction and the average line-of-sight velocity of the features detected towards northeast  is red-shifted  whereas that of the  southwestern features is blue-shifted.
The measured proper motions indicate that the two clusters are moving away from each other along a NE-SW direction with similar velocities, in the range 20--32~\kms.

{ Looking at Fig.~\ref{prmot} (bottom panel), one can note that  the line joining the two maser clusters of Group II does not cross the position of VLA 22~GHz continuum peak, but it is instead shifted towards S-E  by $\gtrsim $0\pas4.

\begin{figure*}
\centering
\includegraphics[width=16cm]{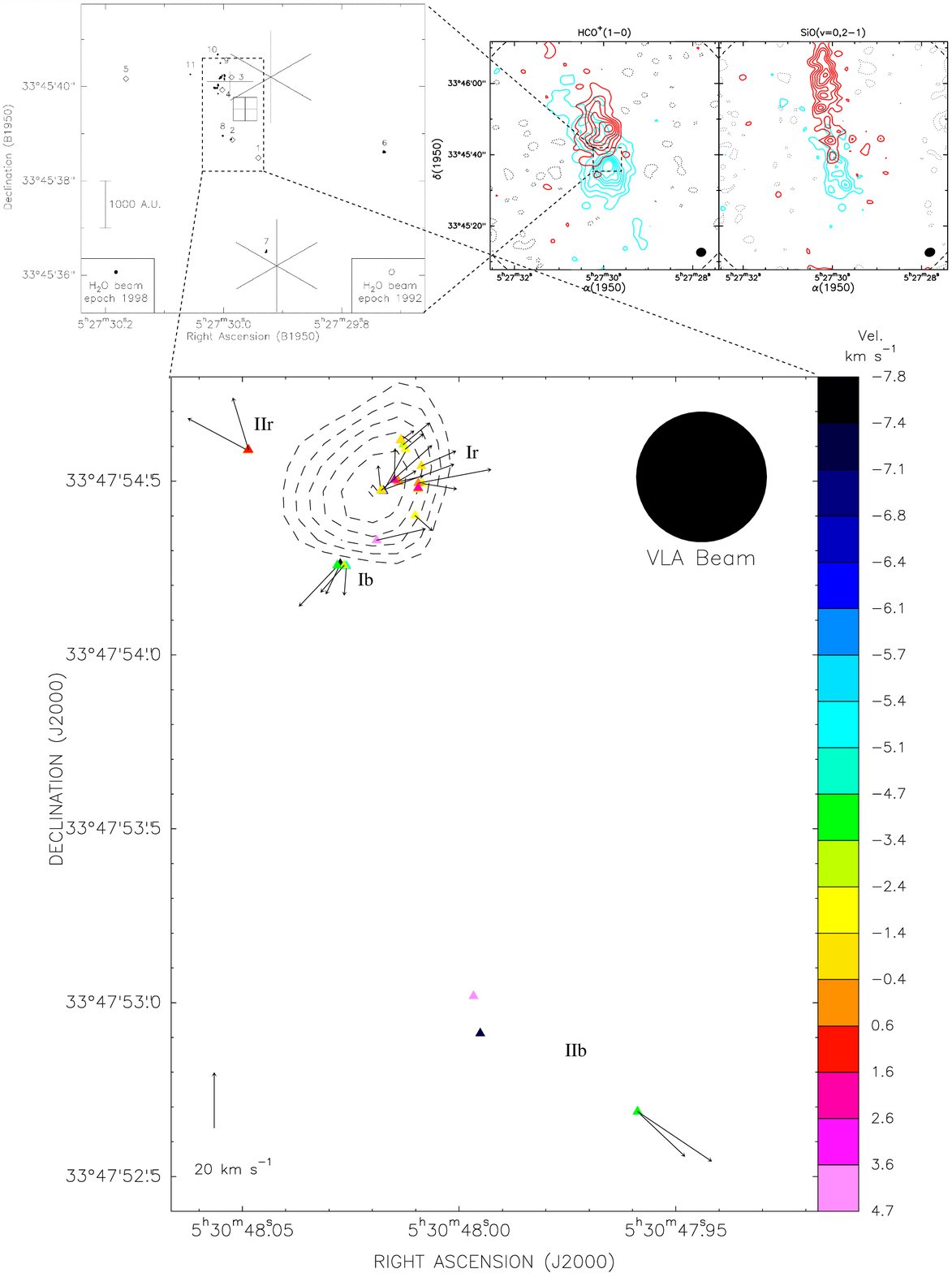}
\caption[VLBA results towards AFGL~5142.]{\scriptsize VLA and VLBA results towards AFGL~5142.\ 
({\itshape Upper right-hand panel}) Contour maps of the OVRO high-velocity emission of HCO$^+$ (1 $\to$ 0) ({\em left}) and SiO ($v = 0$, 2 $\to$ 1) ({\em right}) \citep{Hun99}. 
({\itshape Upper left-hand panel}) Positions  of the 22~GHz masers detected with the VLA in 1992 ({\itshape open squares}) and 1998 ({\itshape filled dots});  OVRO 88~GHz  ({\em boxed  cross}) and  VLA 8.4~GHz ({\em plain cross}) continuum sources \citep{Hun95,Hun99}. 
({\itshape Bottom panel}) Positions of the VLBA H$_2$O maser features are marked with {\it filled triangles}, different {\em colours} denote the features' LSR  velocities, according to the colour scale on the right-hand side of the panel, and the {\em arrows} indicate the measured {\it absolute} proper motions, whose amplitude scale is given at the bottom of the panel; the labels Ib, Ir and IIb, IIr individuate  the blue-shifted and red-shifted maser clusters of Group I and Group II, respectively; the {\em contour} map (with contours representing 5, 6, 7, 8, and 9 times the 50~$\mu$Jy~beam$^{-1}$ RMS noise) shows the   22~GHz VLA continuum emission;  the {\em filled circle} in the upper right corner represents the  restoring beam of the VLA-B 22~GHz continuum map.}
\label{prmot}
\end{figure*}
%_____________________________________________________________________________________________________________
\section{Discussion}
\label{dis}

%Group II
The comparison of the spatial and velocity distribution of the molecular gas traced by the HCO$^+$ and SiO emission and  the Group II  water masers,   suggests that these latter  may be tracing the inner portion of the large-scale molecular outflow.
This interpretation finds support upon three arguments: the elongated distribution of the masers is almost parallel to the position angle of the HCO$^+$ outflow; the blue-shifted features lie northward and the red-shifted ones lie southward, qualitatively matching the velocity distribution of the large-scale molecular outflow; the orientation of the measured proper motions is approximately aligned with the axis of the molecular outflow.

%Group I
 In  Paper I, basing on low-sensitive EVN data,  the observed velocity pattern of Group I masers has been interpreted in terms of a Keplerian rotation. In order to test the Keplerian hypothesis, we fitted the positions and the 3D {\it absolute}  velocities of  maser features with the  Keplerian disk model described in Paper I (see Appendix~\ref{app} for a  more detailed description of the model formalism). The minimizing algorithm  (Eq.~\ref{chi1} in App.~\ref{app}) converged to a best fit solution.
The geometrical parameters of the best fit disk are found to be in good agreement with those previously derived with the EVN maser data (see Paper I).
Since  the measured {\it relative} velocities are in our case three-five times more accurate than {\it absolute} velocities  (see Sect.~\ref{prop_mot}), we  repeated the fit comparing  model predicted and  measured   {\em relative} (to the reference maser feature)  velocities (Eq.~\ref{chi2} in App.~\ref{app}). In this case, the Keplerian model {\em failed} in finding a minimum,  indicating that the measured velocity field for the Group I masers is actually incompatible  with   Keplerian rotation.
This result demonstrates that  the solution obtained fitting both absolute (VLBA) and relative (EVN) proper motions is a spurious result owing to the large measurement errors.
%%%%%{\it accurate} values of measured velocities are essential to avoid erroneous interpretations of H$_2$O maser kinematics.

The unsuccess of the Keplerian model is not surprising. Looking at the Fig.~\ref{prmot}, one notes that the   motion traced by Group I water masers appears to be expansion from (rather than rotation around)  the putative location of the YSO, as  marked by the peak of the   thermal continuum emission. In particular,  proper motions in both Group I (SE and NW) maser clusters are concordly oriented outward from the expected YSO position, whereas in a Keplerian rotation they should be  directed  both inward and outward, unpreferably tracing both the far and near   side of a disk annulus. 

%cone-outflow solution
To test if Group I masers may trace an outflowing motion, the data were fitted with the conical outflow model described in detail in \citet{God05}. Since the absolute values of the measured velocities  do not present a large variation  (compared with the observational errors -- see Table~\ref{af_vlba}), in the conical outflow model we assumed  the maser features to move with  a constant velocity  (differently from the Hubble motion assumption in \citealt{God05}). The best fit solution was looked for minimizing the $\chi^{2}$ given by the sum of the square differences between the model and measured, {\it relative}, 3D velocities (Eq.~\ref{chi2} in App.~\ref{app}). The fit parameters were searched over a wide range of physically plausible values: the cone vertex (i.e., the YSO location) within a distance of 300~AU from the 22~GHz continuum peak; the cone axis orientation over the full 4$\pi$ solid angle;  the cone aperture within the interval $10-85^{\circ}$; the constant expansion velocity from 1 to 50~\kms. The fit parameters were varied with steps comparable or smaller than the associated fit errors.  The minimization algorithm converged towards the best fit solution, whose main characteristics are hereafter briefly resumed.  The best fit conical jet is found to have a large ($\approx 80^{\circ}$) opening angle.  The best fit cone-axis is predicted to form an  angle of $\approx 30^{\circ}$ with the line-of-sight and to have a position angle projected onto the plane of the sky  of $\approx 28^{\circ}$. The (constant) expansion velocity has an absolute value, $v_0$, of 13~\kms. 

 At this point a fundamental question  arises: what might be the driving source of the observed flow motions of Group I and II masers?

First we consider the scenario where  an unique YSO is  responsible for the radio continuum  emission,  the excitation of both maser groups, and the acceleration of the large scale molecular outflow.
%DISK-wind
Since   Group I individuates  a dense ($n_{\rm H_{2}} > 10^{7}$~cm$^{-3}$ -- \citealt{Eli89,Kau96}) clump of gas  around the  expected YSO position, oriented on the sky at large angle from the axis of the jet/outflow system traced by the Group II maser and (at larger scale) by the  HCO$^+$-SiO emission, it is presented the possibility that Group I masers may originate  in the accretion disk surrounding the YSO.
We speculate that  Group I maser features might emerge from the atmosphere of a flared disk (with a disk scale height $H(r)$ scaling linearly with $r$), in the region where modern MHD models predict the launch of the protostellar jet.
As stressed in Sect.~1, such models can be divided into two categories, X-wind and disk-wind.
In the former model, the jet is expected to originate within 0.1~AU from the YSO and to collimate at radii $\leq 10$~AU, whereas the latter model predicts  the wind to be launched across a significant range of disk radii (from tens of AU up to several hundreds of AU, for high-mass YSOs) and to collimate  along the disk axis at correspondingly larger distances from the YSO. The  detected H$_2$O masers are distributed at distances from the  YSO position in the range 10--600~AU. These large distances are clearly incompatible with the X-wind requirements, but they might still find an explanation within the  disk-wind scenario.  In this view, the  disk-wind model would explain  both Group I and Group II water masers,  these latter tracing the collimated portion of the jet-outflow system at larger distances from the YSO ($> 1000$~AU).

%Multiple star system
In the proposed scenario of a single powering source, one would expect the NE-SW outflow axis to   be centered in between the maser clusters Ir and Ib, i.e. at the position of the continuum source. That  is not the case and this is a clear difficulty for the disk-wind interpretation.

The alternative scenario is the one  where  the water masers in AFGL~5142 trace two physically uncorrelated jets (one directed towards SW-NE and the other one towards SE-NW),  each jet being powered by a distinct YSO. 
%In the following we try to identify the two YSOs.
As already discussed in Sect.~\ref{radio}, the YSO  responsible  for the excitation of both the ionized source  and  the Group I masers is expected to be  massive  ($\geq 10~M_{\odot}$).
The other YSO  should  power  the Group II masers and the HCO$^+$-SiO  outflow, which  contains a mass of 35~M$_{\odot}$ \citep{Hun99}. From excitation and energetic considerations one would expect this  YSO to be also a massive one.
Geometrical considerations also suggest that this YSO may be  located on the axis of the outflow traced by the Group II masers. Along this axis, separated by more than 0\pas4 of the 22~GHz continuum peak, no compact continuum emission was detected by our sensitive VLA observations.
 A possible explanation is that the YSO exciting Group II masers might be found in an evolutionary stage earlier than the YSO powering the Group I masers, when the ionized gas is still too much confined by the dense circumstellar environment to be detectable.

We have reviewed the literature looking for indications of a multiple system of YSOs in the region of detection of the water masers.
 \citet{Hun95} and, very recently, \citet{Che05} performed NIR {\it JHK} and H$_2 \ v = 1-0 \ S(1)$ imaging observations of the molecular cloud core in AFGL~5142, revealing the presence of a cluster of YSOs  distributed over a region of diameter of a few arcminutes.
 H$_2$O masers are found in correspondence of the infrared source NIRS1 of \citet{Che05} (IRS1 in the notation of \citealt{Hun95}). The minimum distance between NIRS1 and another member of the cluster is $\gtrsim 5''$ (see Table~2 of \citealt{Che05}), whereas (considering the diameter of the maser distribution) the two (putative) YSOs powering the two maser groups should be separated by a distance  0\pas4 $\lessapprox  d  \lessapprox 2''$ (corresponding to 700$\lessapprox  d  \lessapprox 3600$~AU). Hence, the available NIR observations, having an angular resolution  $\gtrsim  1''$, do not allow us to identify two distinct powering sources for maser Groups I and II.
Towards AFGL~5142, considering an angular scale larger than that presented in Fig.~\ref{prmot} (top right-hand panels), a complex pattern of molecular outflows has also been found, which might suggest  the presence of  a multiple YSO system: \citet{Hun95} detected two CO outflows, nearly perpendicular to each other, one, directed towards SE-NW, extended on scales of a few arcminutes, the other one, directed towards SW-NE, more compact; \citet{Hun99}  detected a HCO$^+$-SiO outflow elongated in a N-S direction (Fig.~\ref{prmot}, upper right-hand panels); \citet{Che05} found H$_2$ jet-like structures along each of the three molecular outflows.
 \citet{Che05} propose that the SW-NE oriented jet/outflow system is powered by NIRS1, whereas the SE-NW oriented jet/outflow system is driven by NIRS30, $30''$ apart from NIRS1 (corresponding to the IRAS source in the AFGL~5142 SFR). Since NIRS1 and NIRS30 are by now the two unique recognized massive YSOs of the AFGL~5142 cluster \citep{Che05}, the available  molecular observations (with an angular resolution  $\gtrsim  1''$) provide no hints for the presence of two massive YSOs within the region of a few arcsec where  the water maser emission is observed. 

In conclusion,  even if the geometry of the outflows traced by the two detected maser Groups may favour the interpretation of two distinct exciting YSOs, the data presently available do not permit unambiguously to identify more than a single YSO in the region of maser emission. 
 In this respect, the axes of the outflows traced by water masers on scales of hundreds and thousands of AU can be good places where to look for the YSOs and the present water maser observations can drive  future studies of the SFR AFGL~5142.

%______________________________________________________________
\section{Conclusions}
\label{con}

 This article reports the results of VLBA multi-epoch  observations of the 22.2~GHz     water masers   and VLA multi-frequency study of the  continuum emission towards  the SFR AFGL~5142.
Respect to the previous EVN observations, the VLBA  provided a significant improvement in sensitivity (by a factor $\sim10$) and an optimized time separation between consecutive epochs ($\sim$1 month vs 3--6 months), resulting in  a larger number of detected maser features (29 vs 12) and measured proper motions (over three or four epochs, 23 vs 2).
The VLBA observations were performed in phase-reference mode, allowing the determination of {\it absolute} positions and proper motions of the water maser features. 

 The water maser emission is found to originate from two elongated structures, with the measured  proper motions aligned along the structures' elongation axes.
Using the same nomenclature of Paper I, we refer to such structures as Group I and Group II.
 Each group consists of   two  (blue- and red-shifted) clusters of features separated by a few hundreds of AU for Group I and thousands of AU for Group II. 

 We detect a compact radio continuum source at 8.4 and 22~GHz (previously reported at 4.9 and 8.4~GHz) located exactly at the center of the Group I maser distribution, indicating that the source ionizing the gas is also responsible for the excitation of the  water masers. 
Comparing our VLA-B observations with previous VLA-C and VLA-D data, there is indication that the continuum emission is extended on angular scales $>4''$ and is optically thin at frequencies larger than  8.4~GHz. The most likely interpretation is in terms of a young HII region.

 The measured  transverse velocities of maser features indicate that the two clusters of Group II are moving away from each other along the cluster-connecting line. The spatial and velocity distribution suggests that the water masers of Group II trace a  collimated outflow. Since  the axis of the larger-scale HCO$^+$ - SiO outflow is almost parallel with the direction of elongation and motion of the Group II maser features,   these latter likely trace the inner portion of the outflow and  originate in  dense clumps  displaced along the outflow axis.

 Group I maser features individuate  a clump of dense gas  around the  expected YSO position (pinpointed by the VLA continuum) and the measured proper motions appear to trace  expansion. 
If a single YSO were responsible for the excitation of both maser Groups, the proposed disk-wind scenario might explain  the observed flow motion of both Group I masers,  tracing the dense material emerging from  the disk atmosphere, and  the Group II maser jet, corresponding to the  portion of the wind collimated along the disk-axis at larger distances from the YSO.
Since the axes joining the two clusters of maser Groups I and II are not intersecting at the position of the continuum location, we favour the scenario where two distinct  massive YSOs are driving the two maser outflows. Presently, the available observational data, from radio to NIR wavelengths, do not permit to identify unambiguously more than a single massive YSO in the region of the detected maser emission. 
Our detection of the maser outflows can drive future studies aiming to localize the powering YSOs.

\appendix
%\onecolumn
\section{Kinematical model formalism}
\label{app}
 In order to investigate  the kinematical structure traced by Group I water masers, we fitted the positions and the 3D (line-of-sight + transverse) velocities of  maser features   with two simple alternative kinematical models: a Keplerian disk  and a conical outflow. The model fitting was performed using both measured {\em absolute} and {\em relative} velocities.
\subsection{Fit of the absolute velocities}
The best-fit solution is found minimising the $\chi^{2}$ given by the sum of the square differences between the  model and measured,  {\it absolute}, 3D velocities:
\begin{equation}
\label{chi1}
\chi^{2} = \sum_{i = 1}^{nspot} \left[ \begin{array}{c} \frac{v^{i}_{z} - (V^{i}_{z} - V_{cloud})}{\Delta V^{i}_{z}} \end{array} \right]^{2} \; + \; 
  \ \sum_{k} \sum_{j : x,y} \left[ \begin{array}{c} \frac{v^{k}_{j} - V^k_j}{\Delta V^{k}_{j}} \end{array} \right]^{2}
\end{equation}
where the lowercase  $v$  denotes the velocity component computed with the model and the uppercase  $V$ is the measured velocity component, with the corresponding uncertainty given by $\Delta V$.
In the first term of the sum,  only the line-of-sight velocities (along the "z" axis) of the identified features (indicated with the index $i$) are considered. 
In order to compare with the model, the measured line-of-sight velocities have been first corrected by the peculiar motion of the  SFR ($V_{\rm cloud} = -4.4$~\kms, as deduced by  CO emission - \citealt{Sne88}), implicitly assuming the LSR  velocity of the (proto-)star to be the same of the surrounding molecular cloud.
The uncertainty of the line-of-sight velocities is taken equal to the spectral FWHM of the maser line, $\lessapprox$ 1~km~s$^{-1}$.
The second term of the sum includes  the  features (indicated with the index $k$) with   measured  transversal velocities, whose components along the R.A. ("x") and DEC ("y") axes are denoted with the index $j$. In the model, the transversal velocity of the YSO is taken to be zero, so that we compared directly the model and the measured {\em absolute} transversal velocities.
\subsection{Fit of the relative velocities}
Fitting absolute velocities has two disadvantages: 1) the absolute motions have  large errors (see Col.~8 of Table~\ref{af_vlba}); 2) the model  assumption that the YSO does not move across the plane of the sky may be wrong.
In order to overcome these limitations, the fit was repeated comparing  model predicted and measured   {\em relative} (to the reference maser feature)  velocities:
\begin{displaymath}
\chi^{2} = \sum_{i = 1}^{nspot} \left[ \begin{array}{c} \frac{(v^{i}_{z} - v^{r}_{z})-(V^{i}_{z} - V^{r}_{z})}{\Delta V^{i}_{z}} \end{array} \right]^{2} \; + \; 
\end{displaymath}
\begin{equation}
\label{chi2}
 + \ \sum_{k} \sum_{j : x,y} \left[ \begin{array}{c} \frac{(v^{k}_{j} - v^{r}_{j}) - V^{k}_{j}}{\Delta V^{k}_{j}} \end{array} \right]^{2}
\end{equation}
where  the  \ $V^{k}_{j}$ are the measured {\em relative} proper motions, $\Delta V^{k}_{j}$ are the associated uncertainties (within a few \kms in all cases) and the index $r$ indicates the reference maser feature. 
 Comparing the model and observed   {\em relative}  velocities has the advantage not to require to include as a further model free parameter the velocity of the YSO.
 
\begin{acknowledgements}
  
We thanks the anonymous referee for pointing to us the presence of previous observations indicating the extended nature of the continuum emission. 

\end{acknowledgements}

\bibliography{biblio}   % bibliography data in biblio.bib

\begin{thebibliography}{23}
\expandafter\ifx\csname natexlab\endcsname\relax\def\natexlab#1{#1}\fi

\bibitem[{{Bonnell} \& {Bate}(2002)}]{Bon02}
{Bonnell}, I.~A. \& {Bate}, M.~R. 2002, \mnras, 336, 659

\bibitem[{{Carpenter} {et~al.}(1990){Carpenter}, {Snell}, \&
  {Schloerb}}]{Car90}
{Carpenter}, J.~M., {Snell}, R.~L., \& {Schloerb}, F.~P. 1990, \apj, 362, 147

\bibitem[{{Carral} {et~al.}(1999){Carral}, {Kurtz}, {Rodr{\'{\i}}guez},
  {Mart{\'{\i}}}, {Lizano}, \& {Osorio}}]{Car99}
{Carral}, P., {Kurtz}, S., {Rodr{\'{\i}}guez}, L.~F., {et~al.} 1999, Revista
  Mexicana de Astronomia y Astrofisica, 35, 97

\bibitem[{{Chen} {et~al.}(2005){Chen}, {Yao}, {Yang}, {Zeng}, \&
  {Sato}}]{Che05}
{Chen}, Y., {Yao}, Y., {Yang}, J., {Zeng}, Q., \& {Sato}, S. 2005, \apj, 629,
  288

\bibitem[{{Elitzur} {et~al.}(1989){Elitzur}, {Hollenbach}, \& {McKee}}]{Eli89}
{Elitzur}, M., {Hollenbach}, D.~J., \& {McKee}, C.~F. 1989, \apj, 346, 983

\bibitem[{{Goddi} {et~al.}(2004){Goddi}, {Moscadelli}, {Alef}, \&
  {Brand}}]{God04}
{Goddi}, C., {Moscadelli}, L., {Alef}, W., \& {Brand}, J. 2004, \aap, 420, 929

\bibitem[{{Goddi} {et~al.}(2005){Goddi}, {Moscadelli}, {Alef}, {Tarchi},
  {Brand}, \& {Pani}}]{God05}
{Goddi}, C., {Moscadelli}, L., {Alef}, W., {et~al.} 2005, \aap, 432, 161

\bibitem[{Hunter {et~al.}(1995)Hunter, Testi, Taylor, Tofani, Felli, \&
  Phillips}]{Hun95}
Hunter, T., Testi, L., Taylor, G., {et~al.} 1995, A\&A, 302, 249

\bibitem[{Hunter {et~al.}(1999)Hunter, Testi, Zhang, \& Sridharan}]{Hun99}
Hunter, T., Testi, L., Zhang, Q., \& Sridharan, T. 1999, AJ, 118, 477

\bibitem[{Imai {et~al.}(2000)Imai, Kameya, Sasao, Miyoshi, Deguchi, Horiuchi,
  \& Asaki}]{Ima00}
Imai, H., Kameya, O., Sasao, T., {et~al.} 2000, ApJ, 538, 751

\bibitem[{{Kaufman} \& {Neufeld}(1996)}]{Kau96}
{Kaufman}, M.~J. \& {Neufeld}, D.~A. 1996, \apj, 456, 250

\bibitem[{{Konigl} \& {Pudritz}(2000)}]{Kon00}
{Konigl}, A. \& {Pudritz}, R.~E. 2000, Protostars and Planets IV, 759

\bibitem[{{McCutcheon} {et~al.}(1991){McCutcheon}, {Dewdney}, {Purton}, \&
  {Sato}}]{McC91}
{McCutcheon}, W.~H., {Dewdney}, P.~E., {Purton}, C.~R., \& {Sato}, T. 1991,
  \aj, 101, 1435

\bibitem[{{Moscadelli} {et~al.}(2005){Moscadelli}, {Cesaroni}, \&
  {Rioja}}]{Mos05}
{Moscadelli}, L., {Cesaroni}, R., \& {Rioja}, M.~J. 2005, \aap, 438, 889

\bibitem[{{Palla} {et~al.}(2002){Palla}, {Zinnecker}, {Maeder}, \&
  {Meynet}}]{Pal02}
{Palla}, F., {Zinnecker}, H., {Maeder}, A., \& {Meynet}, G., eds. 2002,
  {Physics of star formation in galaxies}

\bibitem[{Seth {et~al.}(2002)Seth, Greenhill, \& Holder}]{Set02}
Seth, A., Greenhill, L.~J., \& Holder, B.~P. 2002, ApJ, 581, 325

\bibitem[{{Shu} {et~al.}(1987){Shu}, {Adams}, \& {Lizano}}]{Shu87}
{Shu}, F.~H., {Adams}, F.~C., \& {Lizano}, S. 1987, \araa, 25, 23

\bibitem[{{Shu} {et~al.}(2000){Shu}, {Najita}, {Shang}, \& {Li}}]{Shu00}
{Shu}, F.~H., {Najita}, J.~R., {Shang}, H., \& {Li}, Z.-Y. 2000, Protostars and
  Planets IV, 789

\bibitem[{Snell {et~al.}(1988)Snell, Huang, Dickman, \& Claussen}]{Sne88}
Snell, R.~L., Huang, Y.-L., Dickman, R.~L., \& Claussen, M.~J. 1988, ApJ, 325,
  853

\bibitem[{{Torrelles} {et~al.}(1992){Torrelles}, {Gomez}, {Anglada},
  {Estalella}, {Mauersberger}, \& {Eiroa}}]{Tor92}
{Torrelles}, J.~M., {Gomez}, J.~F., {Anglada}, G., {et~al.} 1992, \apj, 392,
  616

\bibitem[{{Torrelles} {et~al.}(2003){Torrelles}, {Patel}, {Anglada}, {G{\'
  o}mez}, {Ho}, {Lara}, {Alberdi}, {Cant{\' o}}, {Curiel}, {Garay}, \&
  {Rodr{\'{\i}}guez}}]{Tor03}
{Torrelles}, J.~M., {Patel}, N.~A., {Anglada}, G., {et~al.} 2003, \apjl, 598,
  L115

\bibitem[{Vacca {et~al.}(1996)Vacca, Garmany, \& Shull}]{Vac96}
Vacca, W.~D., Garmany, C.~D., \& Shull, J.~M. 1996, ApJ, 460, 914

\bibitem[{{Yorke} \& {Sonnhalter}(2002)}]{Yor02}
{Yorke}, H.~W. \& {Sonnhalter}, C. 2002, \apj, 569, 846

\end{thebibliography}
\bibliographystyle{aa}
\end{document}